\documentclass[twoside,a4paper,11pt]{sea22}
\usepackage{graphicx}
\usepackage{hyperref}
\usepackage{movie15}
\usepackage{color}
\usepackage{booktabs} 
\usepackage{siunitx}  
\usepackage{xcolor}
\usepackage{natbib}    
\usepackage[export]{adjustbox}
\usepackage{pdflscape}
\def\hoy{\number\day \space de \space\ifcase\month\or
 Enero\or Febrero\or Marzo\or Abril\or Mayo\or Junio\or
 Julio\or Agosto\or Septiembre\or Octubre\or Noviembre\or Diciembre\fi
 \space de \number\year}
\def\ii/{\'{\i}}
\def\cion/{ci\'on}
\def\cao/{\c c\~ao}
\def\utw{\smash{\rlap{\lower5pt\hbox{$\sim$}}}}
\def\udtw{\smash{\rlap{\lower6pt\hbox{$\approx$}}}}

\def\tens#1{\ifmmode\mathchoice{\mbox{$\sf\displaystyle#1$}}
{\mbox{$\sf\textstyle#1$}}
{\mbox{$\sf\scriptstyle#1$}}
{\mbox{$\sf\scriptscriptstyle#1$}}\else
\hbox{$\sf\textstyle#1$}\fi}
\def\vec#1{\ifmmode\mathchoice{\mbox{\boldmath$\displaystyle#1$}}
{\mbox{\boldmath$\textstyle#1$}}
{\mbox{\boldmath$\scriptstyle#1$}}
{\mbox{\boldmath$\scriptscriptstyle#1$}}\else
\hbox{\boldmath$\textstyle#1$}\fi}
\def\bbbc{{\mathchoice {\setbox0=\hbox{$\displaystyle\rm C$}\hbox{\hbox
to0pt{\kern0.4\wd0\vrule height0.9\ht0\hss}\box0}}
{\setbox0=\hbox{$\textstyle\rm C$}\hbox{\hbox
to0pt{\kern0.4\wd0\vrule height0.9\ht0\hss}\box0}}
{\setbox0=\hbox{$\scriptstyle\rm C$}\hbox{\hbox
to0pt{\kern0.4\wd0\vrule height0.9\ht0\hss}\box0}}
{\setbox0=\hbox{$\scriptscriptstyle\rm C$}\hbox{\hbox
to0pt{\kern0.4\wd0\vrule height0.9\ht0\hss}\box0}}}}
\def\bbbq{{\mathchoice {\setbox0=\hbox{$\displaystyle\rm
Q$}\hbox{\raise
0.15\ht0\hbox to0pt{\kern0.4\wd0\vrule height0.8\ht0\hss}\box0}}
{\setbox0=\hbox{$\textstyle\rm Q$}\hbox{\raise
0.15\ht0\hbox to0pt{\kern0.4\wd0\vrule height0.8\ht0\hss}\box0}}
{\setbox0=\hbox{$\scriptstyle\rm Q$}\hbox{\raise
0.15\ht0\hbox to0pt{\kern0.4\wd0\vrule height0.7\ht0\hss}\box0}}
{\setbox0=\hbox{$\scriptscriptstyle\rm Q$}\hbox{\raise
0.15\ht0\hbox to0pt{\kern0.4\wd0\vrule height0.7\ht0\hss}\box0}}}}
\def\bbbt{{\mathchoice {\setbox0=\hbox{$\displaystyle\rm
T$}\hbox{\hbox to0pt{\kern0.3\wd0\vrule height0.9\ht0\hss}\box0}}
{\setbox0=\hbox{$\textstyle\rm T$}\hbox{\hbox
to0pt{\kern0.3\wd0\vrule height0.9\ht0\hss}\box0}}
{\setbox0=\hbox{$\scriptstyle\rm T$}\hbox{\hbox
to0pt{\kern0.3\wd0\vrule height0.9\ht0\hss}\box0}}
{\setbox0=\hbox{$\scriptscriptstyle\rm T$}\hbox{\hbox
to0pt{\kern0.3\wd0\vrule height0.9\ht0\hss}\box0}}}}
\def\bbbs{{\mathchoice
{\setbox0=\hbox{$\displaystyle     \rm S$}\hbox{\raise0.5\ht0\hbox
to0pt{\kern0.35\wd0\vrule height0.45\ht0\hss}\hbox
to0pt{\kern0.55\wd0\vrule height0.5\ht0\hss}\box0}}
{\setbox0=\hbox{$\textstyle        \rm S$}\hbox{\raise0.5\ht0\hbox
to0pt{\kern0.35\wd0\vrule height0.45\ht0\hss}\hbox
to0pt{\kern0.55\wd0\vrule height0.5\ht0\hss}\box0}}
{\setbox0=\hbox{$\scriptstyle      \rm S$}\hbox{\raise0.5\ht0\hbox
to0pt{\kern0.35\wd0\vrule height0.45\ht0\hss}\raise0.05\ht0\hbox
to0pt{\kern0.5\wd0\vrule height0.45\ht0\hss}\box0}}
{\setbox0=\hbox{$\scriptscriptstyle\rm S$}\hbox{\raise0.5\ht0\hbox
to0pt{\kern0.4\wd0\vrule height0.45\ht0\hss}\raise0.05\ht0\hbox
to0pt{\kern0.55\wd0\vrule height0.45\ht0\hss}\box0}}}}
\def\bbbz{{\mathchoice {\hbox{$\sf\textstyle Z\kern-0.4em Z$}}
{\hbox{$\sf\textstyle Z\kern-0.4em Z$}}
{\hbox{$\sf\scriptstyle Z\kern-0.3em Z$}}
{\hbox{$\sf\scriptscriptstyle Z\kern-0.2em Z$}}}}
\def\diameter{{\ifmmode\mathchoice
{\ooalign{\hfil\hbox{$\displaystyle/$}\hfil\crcr
{\hbox{$\displaystyle\mathchar"20D$}}}}
{\ooalign{\hfil\hbox{$\textstyle/$}\hfil\crcr
{\hbox{$\textstyle\mathchar"20D$}}}}
{\ooalign{\hfil\hbox{$\scriptstyle/$}\hfil\crcr
{\hbox{$\scriptstyle\mathchar"20D$}}}}
{\ooalign{\hfil\hbox{$\scriptscriptstyle/$}\hfil\crcr
{\hbox{$\scriptscriptstyle\mathchar"20D$}}}}
\else{\ooalign{\hfil/\hfil\crcr\mathhexbox20D}}%
\fi}}
\def\sq{\ifmmode\squareforqed\else{\unskip\nobreak\hfil
\penalty50\hskip1em\null\nobreak\hfil\squareforqed
\parfillskip=0pt\finalhyphendemerits=0\endgraf}\fi}
\def\squareforqed{\hbox{\rlap{$\sqcap$}$\sqcup$}}
\input{isolatin.sty}

\newcommand{\mcii}[1]{\multicolumn{2}{c}{#1}}

\newcommand{\mciv}[1]{\multicolumn{4}{c}{#1}}

\newcommand{\EBV}{\mbox{$E(4405-5495)$}}
\newcommand{\RV}{\mbox{$R_{5495}$}}
\newcommand{\AV}{\mbox{$A_{5495}$}}

\newcommand{\ion}[2]{#1$\,${\sc #2}}
\newcommand{\EWDIBA}{\mbox{EW$_{9366}$}}
\newcommand{\EWDIBB}{\mbox{EW$_{9429}$}}
\newcommand{\EWDIBC}{\mbox{EW$_{9577}$}}
\newcommand{\EWDIBD}{\mbox{EW$_{9632}$}}
\newcommand{\EWCC}{\mbox{EW$_{\rm C_2}$}}
\newcommand{\EWKI}{\mbox{EW$_{\rm K\,I}$}}
\begin{document}
\pagenumbering{arabic}
\pagestyle{myheadings}
\thispagestyle{empty}
\vspace*{-4.0cm}
\textit{\flushleft\small
Highlights of Spanish Astrophysics XIII\\
Proceedings of the XVII Scientific Meeting of the Spanish Astronomical Society \\
\hspace{5mm}held on 13-17 July 2026, in Tarragona, Spain.}
\vspace*{1.0cm}


\begin{flushleft}
{\bf {\LARGE
%
The World Cup in space: Finding ionised buckyballs in the ISM with CARMENES
%
}\\
\vspace*{1cm}
%
J. Maíz Apellániz
%
}\\
\vspace*{0.5cm}
%
Centro de Astrobiolog{\'\i}a, CSIC-INTA, Spain\\
%
\end{flushleft}
%
\markboth{
The World Cup in space
}{ 
%
Ma{\'\i}z Apell\'aniz
%
}
\thispagestyle{empty}
\vspace*{0.4cm}
\begin{minipage}[l]{0.09\textwidth}
\ 
\end{minipage}
\begin{minipage}[r]{0.9\textwidth}
\vspace{1cm}
\section*{Abstract}{\small
%
Over a century after the discovery of the first diffuse interstellar bands (DIBs) in absorption in stellar spectra \citep{Hege22}, 
their origin is still mostly unknown. The exception are four DIBs (out of 600+ known) in the 9300-9700 Å region, for which a possible
ascription to C$_{60}^+$ (ionised buckminsterfullerene, shaped liked a football, hence buckyballs) has been made. However, those DIBs
are located in a spectral region heavily contaminated by telluric H$_2$O absorption, hampering their detection and study from the
ground. I present the results of a study with the CARMENES spectrograph that uses a novel technique with multi-epoch 
spectroscopy at different times of the year to shift the relative position of the DIBs and the surrounding telluric lines, hence 
facilitating the analysis of the DIBs. The technique has been applied to 41 sightlines of diverse extinctions and environments, 
with positive detections in all of them. Our results are consistent with three of those DIBs having C$_{60}^+$ as the carrier.
%
\normalsize}
\end{minipage}
%
%


\section{What is this study about?}

\begin{itemize}
 \item We have collected spectra for 41 sightlines with CARMENES \citep{Quiretal14}, which is an extremely stable instrument that is ideal 
       for exploring regions of high telluric contamination, as part of LiLiMaRlin \citep{Maizetal19a}.
 \item The 41 sightlines (Table~\ref{maintable}) cover a large range of ISM conditions in terms of amount and type of extinction and exposure 
       to UV radiation, including the whole range of sightlines from $\sigma$ to $\zeta$ types \citep{Kreletal97,Camietal97}.
 \item Each star was observed at multiple times of the year to shift the relative position of the telluric lines and DIBs and the spectra 
       combined using the UNWIND software \citep{Maizetal26b}.
 \item Each DIB was fitted using the profiles determined by \citet{Maizetal26b} using the ISM velocities measured fron \ion{K}{i} and 
       C$_2$ lines.
 \item For each sightline we measured the EWs of \ion{K}{i}~$\lambda$7698.974 and C$_2$~$\lambda$8763.751 and the extinction parameters 
       using the family of \citet{Maizetal14a}. Whenever available, we used the extinction parameters of 
       \citet{MaizBarb18,Maizetal21a} but for 11 we calculated them anew using CHORIZOS \citep{Maiz04c} and the photometric calibrations for
       \textit{Gaia} of \citet{MaizWeil25} and for 2MASS of \citet{MaizPant18}.
\end{itemize}


\section{Comparison with molecular carbon}      

$\,\!$\indent The left panel of Fig.~\ref{fig2} shows that DIBN9577 is only weakly correlated with C$_2$~$\lambda$8763.751, as opposed to the 
class known as C$_2$ DIBs \citep{Thoretal03}. Hence, DIBN5977 is a non-C$_2$ DIB, making C$_{60}^+$ a potential carrier, which is expected to 
 be more abundant in the UV-exposed regions of the ISM ($\sigma$ sightlines). We use the lines plotted in that panel to divide our 
sightlines in C$_2$ rich (3 targets), normal (27), and C$_2$ poor (5), to which we add the 6 targets with no C$_2$ detection. The three 
C$_2$ rich sightlines correspond to stars located behind dust lanes (W~40~OS~1a, Bajamar star, and NGC 2024-1) at distances shorter than 1~kpc,
thus confirming the association with molecular gas. The 6 sightlines without C$_2$ detections are exposed to UV radiation and include the 
prototypical $\sigma$~Sco~Aa,Ab.


\section{Comparison between the two best DIBs}      

$\,\!$\indent The two strongest and cleanest of our DIBs, DIBN9577 and DIBN9632, are detected in all sightlines but we do not include in 
Table~\ref{maintable} the values for B+A and some late-O stars because they are contaminated by the stellar absorption from the 
\ion{Mg}{ii}~$\lambda$9632.1 triplet \citep{Galaetal21}. In the right panel of Fig.~\ref{fig1} we plot the EWs for the 26 uncontaminated 
sightlines and we find a near-perfect correlation with a Pearson coefficient of 0.979. This value is much higher that the one of 0.37 found 
by \citet{Galaetal21}, indicating that their sample likely had residual stellar contamination. It is also higher than the value of 0.89 found
with the 9-star sample by \citet{Nieetal22} but close to their value of 0.96 once unsertainties are included. Therefore, we agree with the
\citet{Nieetal22} conclusion that both DIBs likely originate in the same carrier.


\section{Comparison with the two other DIBs}      

\begin{landscape}
\begin{figure}
 \vspace{-1cm}
 \centerline{\includegraphics[width=0.55\textwidth]{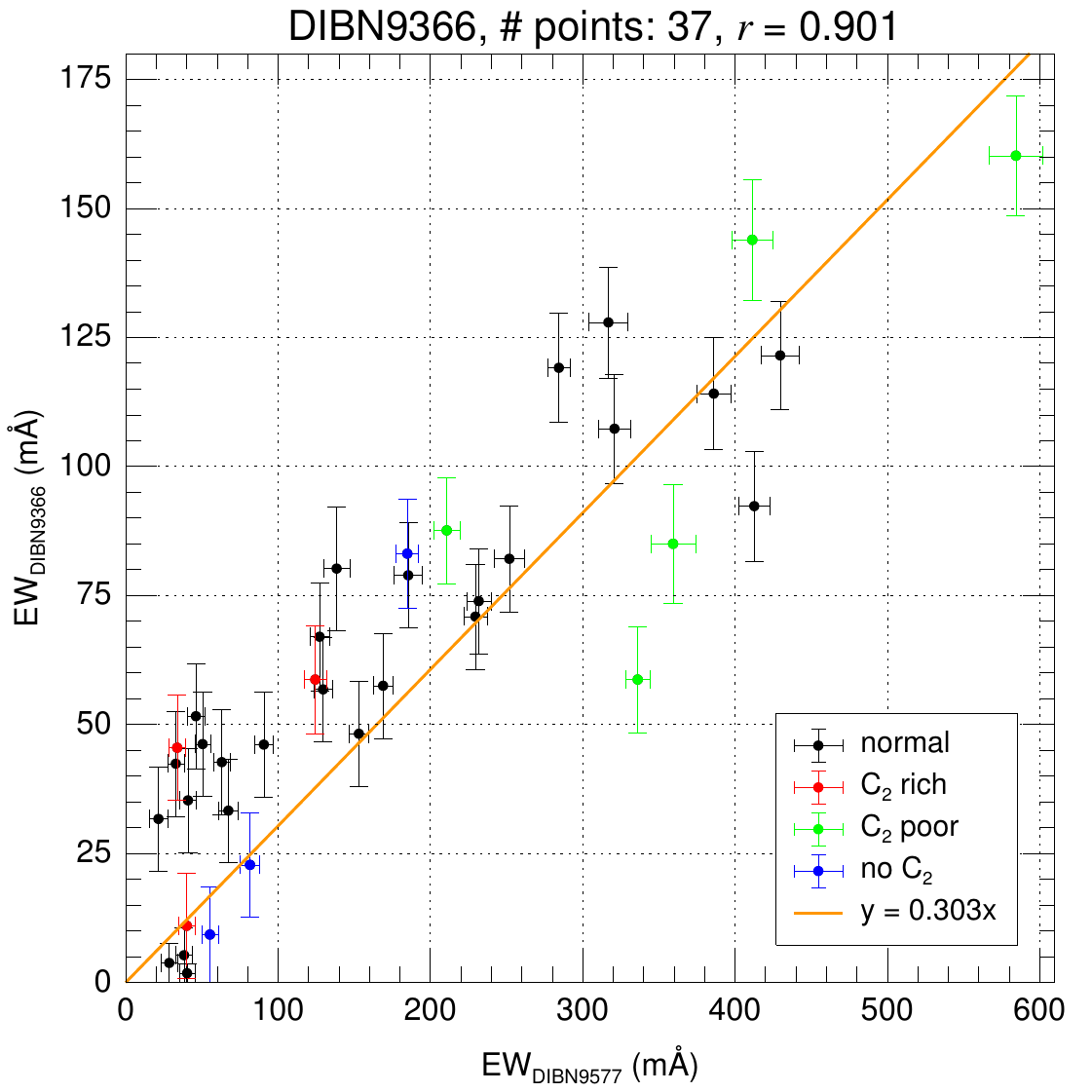} \
             \includegraphics[width=0.55\textwidth]{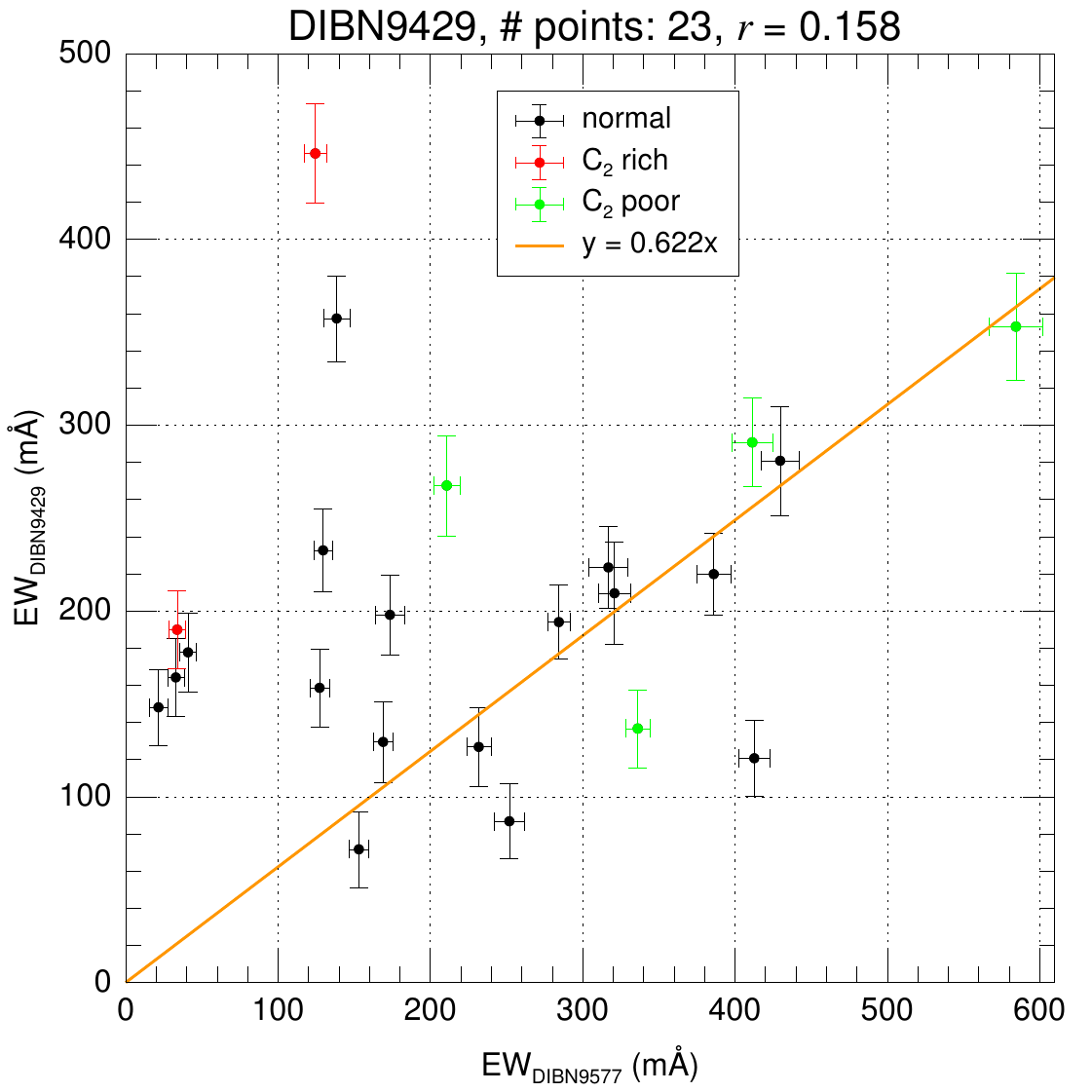} \
             \includegraphics[width=0.55\textwidth]{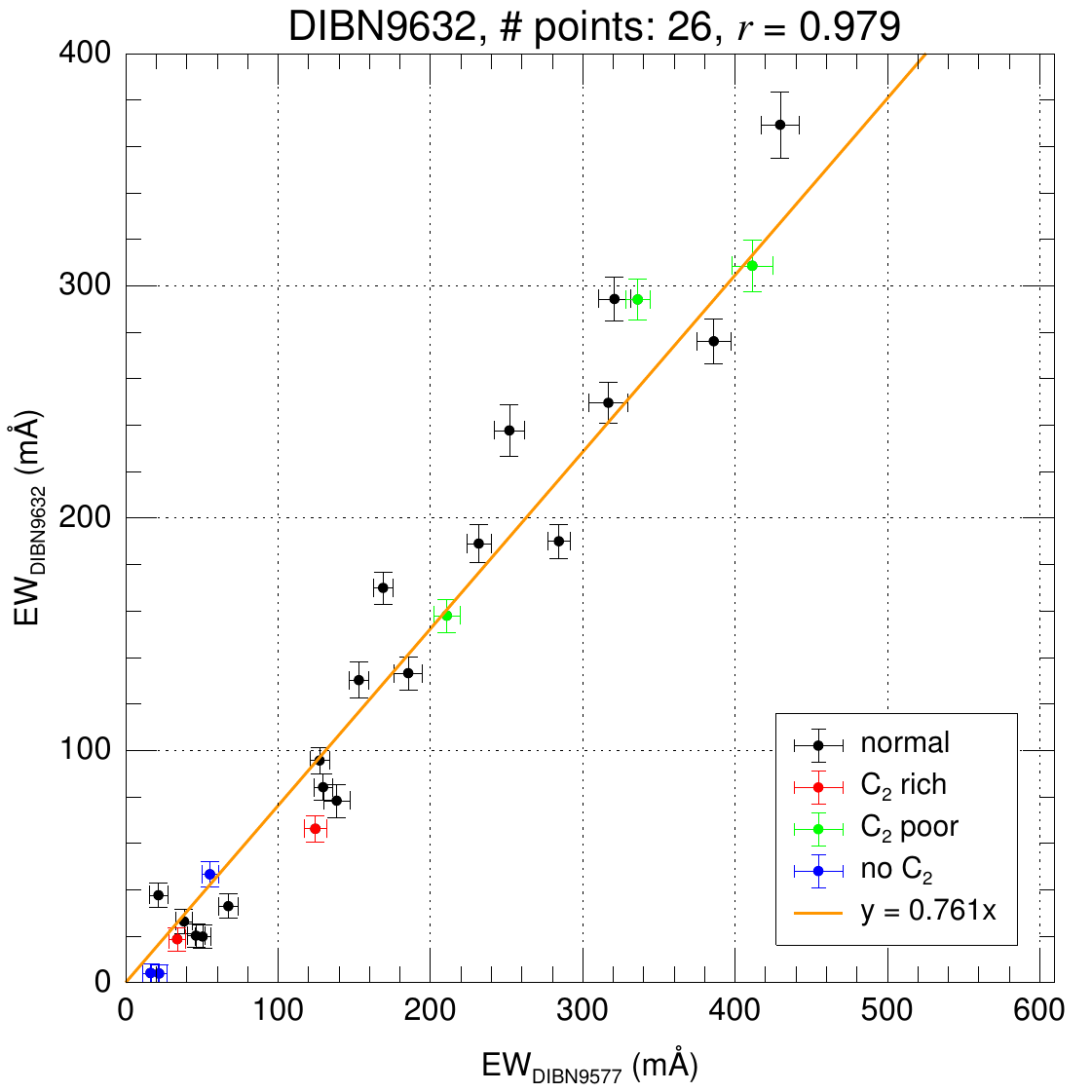}}
 \caption{EW-EW correlation plots comparing DIBN9577 ($x$ axis) with the other three DIBs ($y$ axis). The title gives the number of points in the
          plot and the Pearson correlation coefficient. The data are divided by the C$_2$ strength in the left panel of Fig.~\ref{fig2}. The 
          orange line shows an average constant ratio between the two EWs.}
 \label{fig1}   
\end{figure}
\end{landscape}

$\,\!$\indent DIBN9366 is weaker than DIBN9577 or DIBN9632 and is located in a region with stronger telluric lines but is not affected by a
stellar line like DIBN9632. Our multi-epoch technique allows us to measure its EW in 37 of the 41 sightlines. The left panel of Fig.~\ref{fig1}
shows a strong correlation, with a Pearson coefficient of 0.901. Including the effect of random (and possibly systematic at low values of the EW)
uncertainties would increase that value, so it is possible that it originates in the same carrier as the other two.

DIBN9429 is a relatively strong DIB but is located in the region with the stronger telluric contamination of the four DIBs, so we can only
measure its EW for cases where the DIB is strong and we have data that includes epochs with low H$_2$O columns (23 sightlines). With those
caveats we find that the Pearson coefficient is just 0.158 (center panel of Fig.~\ref{fig1}), making it unlikely that it originates in the same 
carrier.


\section{Comparison with K$\,${\small I} and colour excess}      

\begin{landscape}
\begin{figure}
 \vspace{1cm}
 \centerline{\includegraphics[width=0.55\textwidth]{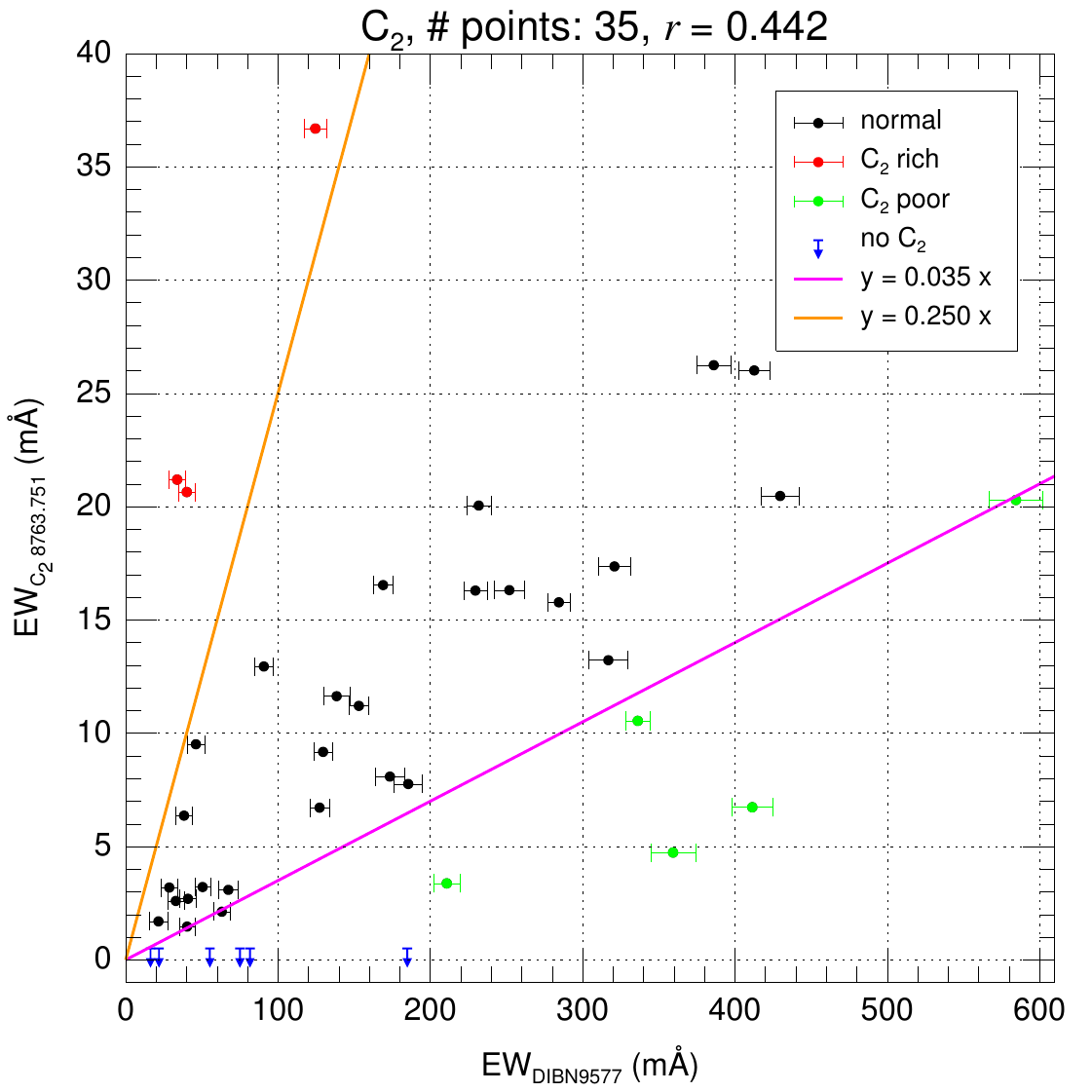} \
             \includegraphics[width=0.55\textwidth]{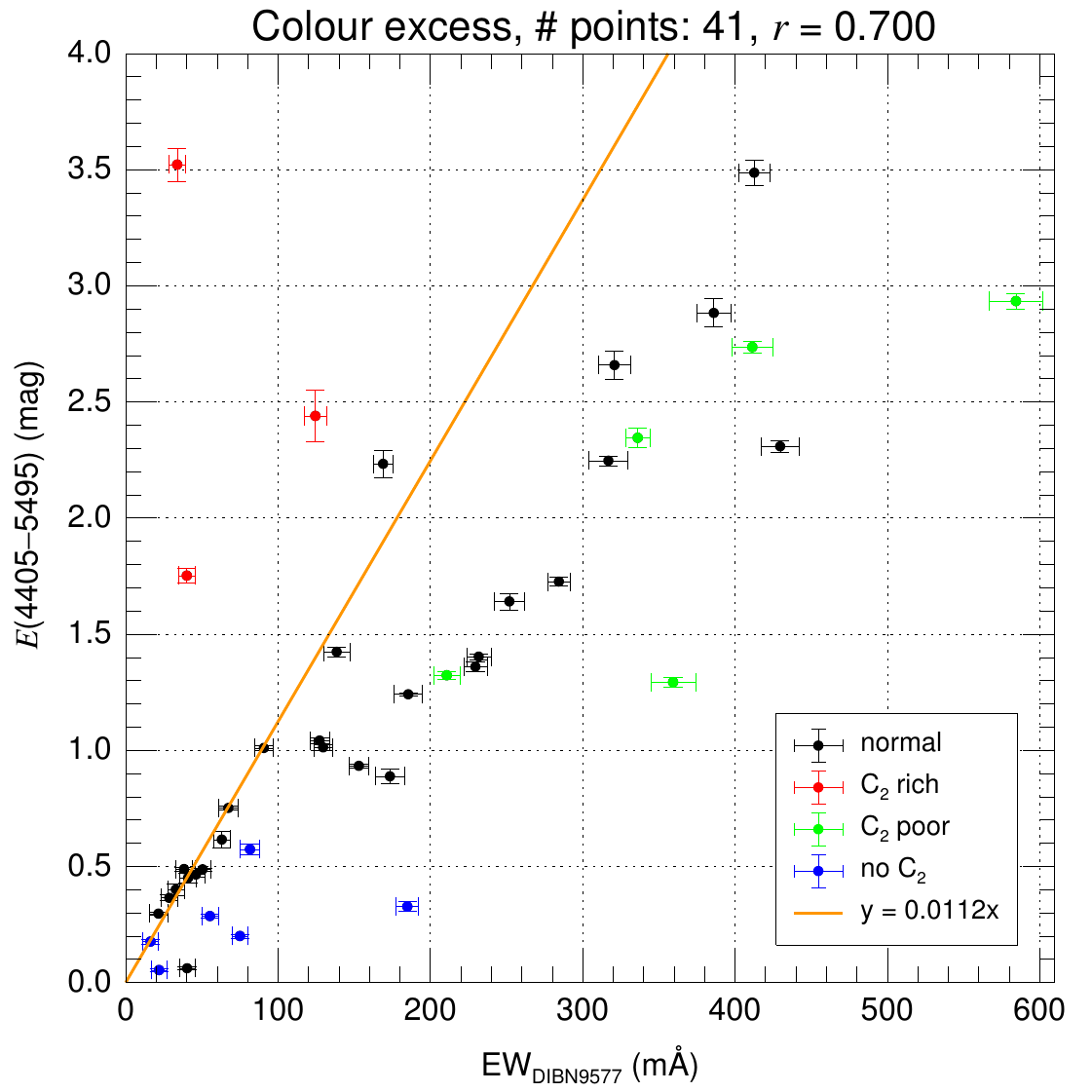} \
             \includegraphics[width=0.55\textwidth]{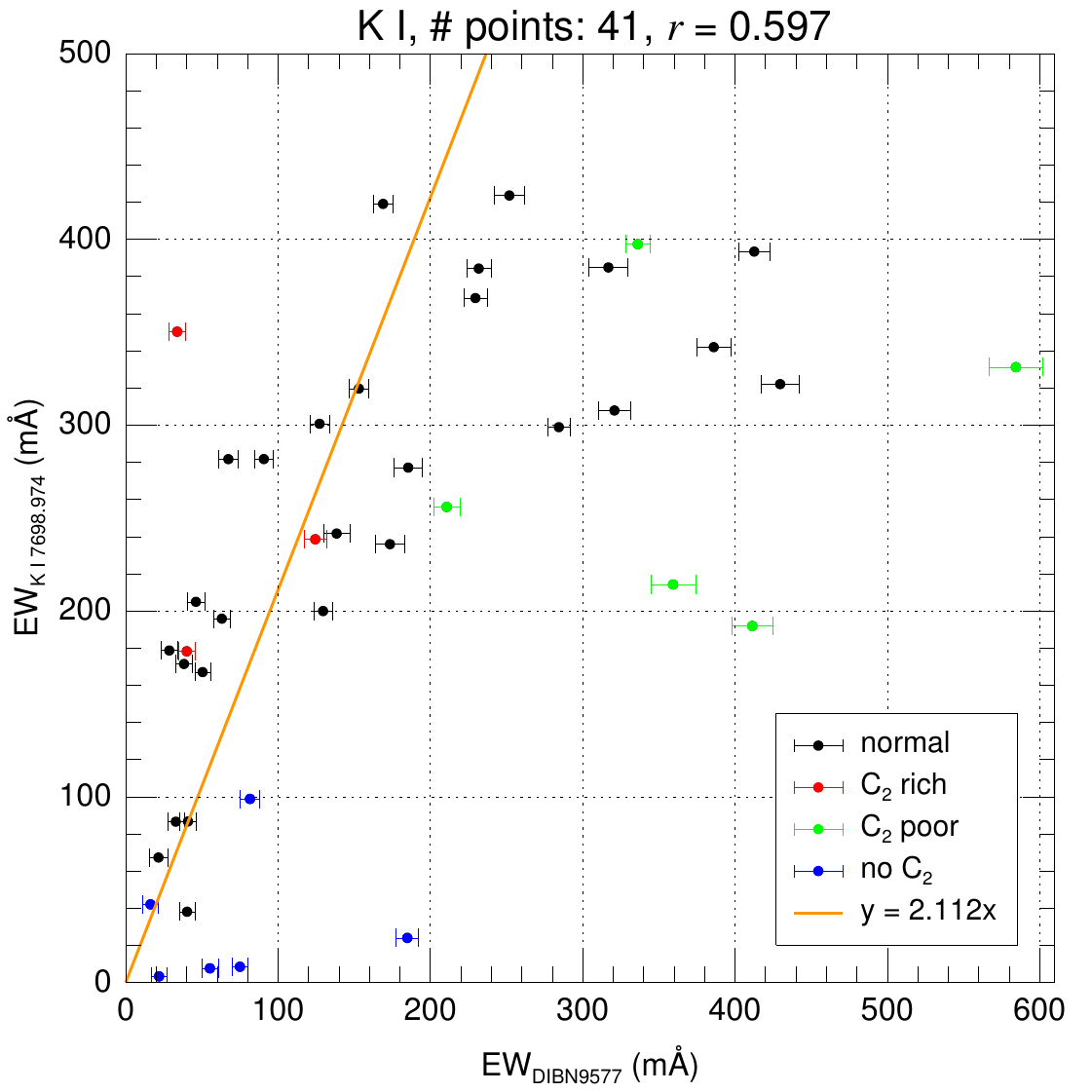}}
 \caption{Correlation plots comparing the EW DIBN9577 in the $x$ axis with the EW of C$_2$~$\lambda$8763.751 (left), the monochromatic colour 
          excess \EBV\ (center) and the EW of \ion{K}{i}~$\lambda$7698.974 (right) in the $y$ axis. The title gives the number of points in the
          plot and the Pearson correlation coefficient. For C$_2$ the data are divided in three regions according to the relative strength of the
          line. In the other two, the orange line shows an average constant ratio between the two quantities.}
 \label{fig2}   
\end{figure}
\end{landscape}

$\,\!$\indent The EW of \ion{K}{i}~$\lambda$7698.974 has a Pearson correlation coefficient of 0.597 with the EW of DIBN9577 (right panel of
Fig.~\ref{fig2}), higher than the comparison with $C_2$ but lower than that of DIB8366 or DIBN9632. However, there are two caveats: the 
\ion{K}{i} profiles are saturated or close to for large values of the EW (see e.g. \citealt{Maizetal21a}) and the worst offenders are the
sightlines that are C$_2$ poor or with no C$_2$ detections. The logical interpretation is that \ion{K}{i} traces an intermediate ISM phase
between the UV-shielded C$_2$ regions and the UV-exposed C$_{60}^+$ regions.

The center panel of Fig.~\ref{fig2} shows the correlation plot between the EW of DIBN9577 and the monochromatic colour excess
\EBV\ \citep{Maizetal14a,Maiz24}. The Pearson correlation coefficient is quite high, 0.700, and the scatter conforms to the same distribution as
in the comparison with the EW of C$_2$ (in terms of C$_2$ richness). The interpretation here is that dust is present in a variety of
environments (UV exposed, intermediate, and shielded) with DIBN9577, \ion{K}{i}~$\lambda$7698.974, and C$_2$~$\lambda$8763.751 being
representative tracers of each one of them.

\section{Conclusions}

\begin{itemize}
 \item DIBN9577 and DIBN9632 (and possibly DIBN9366) quite likely originate in the same carrier.
 \item DIBN9366 appears to have a different origin.
 \item The evidence from different ISM tracers suggests that DIBN9577 is a $\sigma$ DIB (formed in UV exposed regions), which is consistent
       with C$_{60}^+$ being the carrier.
\end{itemize}

\section{References}

\bibliographystyle{aa} 
\bibliography{general}\vspace{0.75in}

\begin{landscape}
\thispagestyle{empty}
\begin{table}
\vspace{-2cm}
\newcommand{\hh}{\hspace{-5mm}}
\caption{Sample information and results.}
\centerline{\small
\begin{tabular}{lllllr@{$\pm$}lr@{$\pm$}lr@{$\pm$}lr@{$\pm$}lr@{$\pm$}lr@{$\pm$}lr@{$\pm$}lrr}
\midrule
Name                      & \mciv{Spectral classification}        & \mcii{\hh\EBV} & \mcii{\RV} & \mcii{\AV}   & \mcii{\EWDIBA} & \mcii{\EWDIBB} & \mcii{\EWDIBC} & \mcii{\EWDIBD} & \EWCC  & \EWKI  \\
                          & ST    & LC   & qual.   & secondary    & \mcii{(mag)}   & \mcii{}    & \mcii{(mag)} & \mcii{(m\AA)}  & \mcii{(m\AA)}  & \mcii{(m\AA)}  & \mcii{(m\AA)}  & (m\AA) & (m\AA) \\
\midrule
$\sigma$~Sco~Aa,Ab        & B0.7  & IV  &          & B1: V        &    0.327&0.020 &  4.33&0.38 &  1.419&0.046 &          83&11 &     \mcii{---} &         185&7  &     \mcii{---} &    --- &   24.0 \\
$\rho$~Oph~A              & B2    & IV  & (n)      &              &    0.450&0.022 &  4.27&0.45 &  1.918&0.113 &          35&10 &         178&21 &          41&5  &     \mcii{---} &    2.7 &   86.7 \\
$\rho$~Oph~B              & B2.5  & IV  & nn       &              &    0.401&0.026 &  5.36&0.62 &  2.147&0.122 &          42&10 &         164&21 &          33&5  &     \mcii{---} &    2.6 &   86.6 \\
$\zeta$~Oph               & O9.2  & IV  & nn(e)    &              &    0.297&0.006 &  3.19&0.10 &  0.945&0.016 &          32&10 &         148&21 &          21&6  &          38&5  &    1.7 &   67.3 \\
GLS~\num{4962}            & ON5   & I   & fp       &              &    1.424&0.021 &  3.23&0.05 &  4.601&0.026 &          80&12 &         357&23 &         139&8  &          78&7  &   11.6 &  241.6 \\
BD~$-$12~4979             & O9.5  & V   &          &              &    0.888&0.031 &  3.43&0.15 &  3.044&0.046 &     \mcii{---} &         198&22 &         174&9  &     \mcii{---} &    8.1 &  235.9 \\
BD~$-$11~4586             & O8    & Ib  & (f)      &              &    1.241&0.007 &  3.42&0.03 &  4.242&0.023 &          79&10 &     \mcii{---} &         186&9  &         133&7  &    7.8 &  277.1 \\
HD~\num{168075}           & O5.5: & V   & ((f))z   & B0: V        &    0.752&0.010 &  3.58&0.07 &  2.692&0.025 &          33&10 &     \mcii{---} &          67&6  &          33&5  &    3.1 &  281.7 \\
GLS~\num{19613}~A         & O2/4  & V   & p        &              &    2.736&0.027 &  3.85&0.03 & 10.532&0.028 &         144&12 &         291&24 &         411&14 &         309&11 &    6.7 &  191.9 \\
W~40~OS~1a                & O9:   & V   &          &              &    2.440&0.111 &  4.47&0.18 & 10.903&0.071 &          59&11 &         446&27 &         125&7  &          66&6  &   36.7 &  238.5 \\
HD~\num{183143}           & B8    & Ia  & e        &              &    1.293&0.023 &  3.21&0.07 &  4.146&0.036 &          85&12 &     \mcii{---} &         359&15 &     \mcii{---} &    4.7 &  214.2 \\
HDE~\num{344777}          & O7.5  & V   &          &              &    1.013&0.011 &  3.18&0.06 &  3.222&0.028 &          57&10 &         233&22 &         130&6  &          84&6  &    9.2 &  199.9 \\
Cyg~X-1                   & O9.7  & Iab & p var    &              &    1.042&0.014 &  3.44&0.06 &  3.585&0.025 &          67&10 &         159&21 &         127&6  &          96&6  &    6.7 &  300.7 \\
25~Peg                    & B7    & IV  & (n)e     &              &    0.061&0.007 &  2.43&0.46 &  0.148&0.018 &           2&2  &     \mcii{---} &          40&5  &     \mcii{---} &    1.5 &   38.1 \\
Cyg~OB2-B17               & O6    & Ia  & f        & O9: Ia:      &    2.883&0.061 &  3.00&0.08 &  8.650&0.084 &         114&11 &         220&22 &         386&11 &         276&10 &   26.2 &  341.9 \\
GLS~\num{15133}           & O9.5  & IV  &          &              &    2.309&0.026 &  3.06&0.03 &  7.062&0.023 &         122&10 &         281&29 &         430&13 &         369&14 &   20.5 &  322.1 \\
Cyg~OB2-A11               & O7    & Ib  & (f)      &              &    2.658&0.059 &  3.15&0.07 &  8.382&0.046 &         107&11 &         210&27 &         321&11 &         294&9  &   17.4 &  307.9 \\
Cyg~OB2-12                & B5    & Ia+ &          &              &    3.486&0.056 &  3.08&0.09 & 10.739&0.204 &          92&11 &         121&21 &         413&10 &     \mcii{---} &   26.0 &  393.4 \\
GLS~\num{15148}           & O6.5: & V   &          &              &    2.247&0.021 &  3.31&0.04 &  7.434&0.032 &         128&11 &         223&22 &         317&13 &         250&9  &   13.2 &  384.8 \\
2MASS~J20395358$+$4222505 & B0    & I   &          &              &    2.934&0.034 &  3.40&0.04 &  9.969&0.036 &         160&12 &         353&29 &         585&18 &     \mcii{---} &   20.3 &  331.2 \\
Bajamar~star              & O3.5  & III & (f*)     & O8:          &    3.522&0.071 &  2.96&0.05 & 10.413&0.051 &          46&10 &         190&21 &          34&5  &          19&5  &   21.2 &  350.3 \\
GLS~\num{11448}           & O3.5  & II  & (f*)     & O3.5 II(f*)  &    1.726&0.019 &  3.23&0.05 &  5.568&0.045 &         119&11 &         194&20 &         284&8  &         190&7  &   15.8 &  298.9 \\
GLS~\num{11449}           & O4.5  & IV  & (f)      &              &    1.323&0.018 &  3.21&0.07 &  4.253&0.038 &          88&10 &         267&27 &         211&9  &         158&7  &    3.4 &  256.0 \\
HD~\num{206267}~Aa,Ab     & O5    & V   & ((fc))   & B0: V + O9 V &    0.465&0.009 &  3.41&0.11 &  1.587&0.025 &          52&10 &     \mcii{---} &          46&6  &          20&5  &    9.5 &  204.9 \\
$\lambda$~Cep             & O6.5  & Iab & (n)fp    &              &    0.487&0.006 &  3.25&0.08 &  1.586&0.028 &          46&10 &     \mcii{---} &          51&5  &          20&5  &    3.2 &  167.1 \\
6~Cas~A                   & A3    & Ia  &          &              &    0.616&0.036 &  3.56&0.21 &  2.182&0.021 &          43&10 &     \mcii{---} &          63&6  &     \mcii{---} &    2.1 &  195.8 \\
BD~$+$66~1661             & O9.5  & IV  &          &              &    1.010&0.012 &  3.12&0.06 &  3.152&0.029 &          46&10 &     \mcii{---} &          91&6  &     \mcii{---} &   12.9 &  281.7 \\
V747~Cep                  & O5.5  & V   & (n)((f)) &              &    1.641&0.035 &  3.13&0.08 &  5.144&0.035 &          82&10 &          87&20 &         252&10 &         238&11 &   16.3 &  423.6 \\
BD~$+$66~1675             & O7.5  & V   & z        & O8 V + B     &    1.403&0.012 &  2.84&0.04 &  3.986&0.027 &          74&10 &         127&21 &         232&8  &         189&8  &   20.0 &  384.3 \\
BD~$+$66~1674             & B0    & V   &          & B0 V         &    1.360&0.022 &  2.97&0.07 &  4.038&0.041 &          71&10 &     \mcii{---} &         230&8  &     \mcii{---} &   16.3 &  368.4 \\
GLS~\num{17957}           & O7    & V   & ((f))z   &              &    2.233&0.058 &  3.29&0.09 &  7.342&0.048 &          58&10 &         130&22 &         169&7  &         170&7  &   16.5 &  419.0 \\
GLS~\num{20775}           & O8    & V   &          &              &    2.345&0.043 &  3.48&0.06 &  8.154&0.038 &          59&10 &         137&21 &         336&8  &         294&9  &   10.5 &  397.3 \\
HD~\num{15570}            & O4    & I   & f        &              &    0.933&0.009 &  3.49&0.05 &  3.259&0.021 &          48&10 &          72&20 &         153&7  &         130&8  &   11.2 &  319.6 \\
HD~\num{21389}            & A0    & Ia  &          &              &    0.573&0.023 &  3.55&0.22 &  2.033&0.080 &          23&10 &     \mcii{---} &          82&6  &     \mcii{---} &    --- &   98.8 \\
AE~Aur                    & O9.5  & V   &          &              &    0.489&0.008 &  3.67&0.09 &  1.795&0.024 &           5&5  &     \mcii{---} &          38&5  &          26&5  &    6.4 &  171.4 \\
3~Gem~A                   & B2.5  & Ib  &          &              &    0.365&0.012 &  3.42&0.16 &  1.246&0.025 &           4&4  &     \mcii{---} &          29&5  &     \mcii{---} &    3.2 &  178.7 \\
$\lambda$~Ori~A           & O8    & III & ((f))    &              &    0.177&0.011 &  4.22&0.41 &  0.745&0.028 &     \mcii{---} &     \mcii{---} &          16&5  &           4&4  &    --- &   42.1 \\
15~Mon~Aa,Ab              & O7    & V   & ((f))z   & B1: Vn       &    0.054&0.006 &  4.43&0.75 &  0.238&0.018 &     \mcii{---} &     \mcii{---} &          22&5  &           4&4  &    --- &    3.3 \\
NGC~2024-1                & O9.7  & V   &          & B1: V        &    1.752&0.033 &  4.72&0.09 &  8.263&0.030 &          11&10 &     \mcii{---} &          40&5  &     \mcii{---} &   20.6 &  178.3 \\
$\theta^1$~Ori~Ca,Cb      & O7    &     & f?p var  &              &    0.286&0.009 &  6.17&0.24 &  1.767&0.026 &           9&9  &     \mcii{---} &          55&5  &          47&5  &    --- &    7.7 \\
$\theta^2$~Ori~A          & O9.2  & V   &          & B0.5: V(n)   &    0.200&0.008 &  6.22&0.33 &  1.246&0.023 &     \mcii{---} &     \mcii{---} &          75&5  &     \mcii{---} &    --- &    8.5 \\
\midrule
\end{tabular}
}
\label{maintable}
\end{table}

\end{landscape}


\end{document}